\documentclass[modern]{aastex62}
\newcommand\msun {M_\odot}

\newcommand\mearth {{M_\oplus}}

\newcommand{\mathbold}[1]{\mbox{\boldmath $\bf#1$}}

\newcommand\ltsima{$\; \buildrel <\over\sim \;$}
\newcommand\simlt{\lower.5ex\hbox{\ltsima}}
\newcommand\gtsima{$\; \buildrel >\over\sim \;$}
\newcommand\simgt{\lower.5ex\hbox{\gtsima}}

\shorttitle{Microlensing vs the Core Accretion Model}
\shortauthors{Suzuki et al.}
\begin{document}

\title{Microlensing Results Challenge the Core Accretion Runaway Growth Scenario for Gas Giants}

\correspondingauthor{Daisuke Suzuki}
\email{dsuzuki@ir.isas.jaxa.jp}

\author[0000-0002-5843-9433]{Daisuke Suzuki}
\affil{Institute of Space and Astronautical Science, Japan Aerospace Exploration Agency, 3-1-1 Yoshinodai, Chuo, Sagamihara, Kanagawa 252-5210, Japan
}

\author[0000-0001-8043-8413]{David P. Bennett}
\affiliation{Laboratory for Exoplanets and Stellar Astrophysics, NASA Goddard Space Flight Center, Greenbelt, MD 20771, USA}
\affiliation{Department of Astronomy, University of Maryland, College Park, MD 20742, USA}
%\affiliation{}

\author{Shigeru Ida}
\affiliation{Earth-Life Science Institute, Tokyo Institute of Technology, Meguro-ku, Tokyo, 152-8550, Japan}

\author[0000-0002-1013-2811]{Christoph Mordasini}
\affiliation{Univerisity of Bern, Physikalisches Institut, Gesellschaftsstrasse 6, CH-3012, Bern, Switzerland}

\author{Aparna Bhattacharya}
\affiliation{Laboratory for Exoplanets and Stellar Astrophysics, NASA Goddard Space Flight Center, Greenbelt, MD 20771, USA}
\affiliation{Department of Astronomy, University of Maryland, College Park, MD 20742, USA}

\author{Ian A. Bond}
\affiliation{Institute of Information and Mathematical Sciences, Massey university, Private Bag 102-904, North Shore Mail Centre, Auckland, New Zealand}

\author{Martin Donachie}
\affiliation{Department of Physics, University of Auckland, Private Bag 92019, Auckland, New Zealand}

\author[0000-0002-4909-5763]{Akihiko Fukui}
\affiliation{Department of Earth and Planetary Science, The University of Tokyo, 7-3-1 Hongo, Bunkyo-ku, Tokyo 113-0033, Japan}
\affiliation{Instituto de Astrof\'isica de Canarias, V\'ia L\'actea s/n, E-38205 La Laguna, Tenerife, Spain}

\author{Yuki Hirao}
\affiliation{Laboratory for Exoplanets and Stellar Astrophysics, NASA Goddard Space Flight Center, Greenbelt, MD 20771, USA}
\affiliation{Department of Earth and Space Science, Graduate School of Science, Osaka University, 1-1 Machikaneyama, Toyonaka, Osaka 560-0043, Japan}

\author[0000-0003-2302-9562]{Naoki Koshimoto}
\affiliation{Department of Astronomy, Graduate School of Science, The University of Tokyo, 7-3-1 Hongo, Bunkyo-ku, Tokyo 113-0033, Japan}
\affiliation{National Astronomical Observatory of Japan, 2-21-1 Osawa, Mitaka, Tokyo 181-8588, Japan}

\author[0000-0001-9818-1513]{Shota Miyazaki}
\affiliation{Department of Earth and Space Science, Graduate School of Science, Osaka University, 1-1 Machikaneyama, Toyonaka, Osaka 560-0043, Japan}

\author{Masayuki Nagakane}
\affiliation{Department of Earth and Space Science, Graduate School of Science, Osaka University, 1-1 Machikaneyama, Toyonaka, Osaka 560-0043, Japan}

\author{Cl\'ement Ranc}
\affiliation{Laboratory for Exoplanets and Stellar Astrophysics, NASA Goddard Space Flight Center, Greenbelt, MD 20771, USA}

\author[0000-0001-5069-319X]{Nicholas J. Rattenbury}
\affiliation{Department of Physics, University of Auckland, Private Bag 92019, Auckland, New Zealand}

\author{Takahiro Sumi}
\affiliation{Department of Earth and Space Science, Graduate School of Science, Osaka University, 1-1 Machikaneyama, Toyonaka, Osaka 560-0043, Japan}

\author{Yann Alibert}
\affiliation{University of Bern, Physikalisches Institut, Gesellschaftsstrasse 6, CH-3012 Bern, Switzerland}

\author{Douglas N.C. Lin}
\affiliation{UCO/Lick Observatory, Board of Studies in Astronomy and Astrophysics, University of California, Santa Cruz, CA 95064, USA}
\affiliation{Institute for Advanced Studies, Tsinghua University, Beijing, China}



\begin{abstract}

We compare the planet-to-star mass-ratio distribution measured by gravitational 
microlensing to core accretion theory predictions from population synthesis models. 
The core accretion theory's runaway gas accretion process predicts a dearth of 
intermediate-mass giant planets that is not seen in the microlensing results. 
In particular, the models predict $\sim10\,\times$ fewer planets at mass ratios of 
$10^{-4} \leq q \leq 4 \times 10^{-4}$ than inferred from microlensing observations. 
This tension implies that gas giant formation may involve processes that have hitherto 
been overlooked by existing core accretion models or that the planet-forming environment 
varies considerably as a function of host-star mass. Variation from the usual assumptions 
for the protoplanetary disk viscosity and thickness could reduce this discrepancy, but 
such changes might conflict with microlensing results at larger or smaller mass ratios, or
with other observations. The resolution of this discrepancy 
may have important implications for planetary habitability because it has been suggested 
that the runaway gas accretion process may have triggered the delivery of water to 
our inner solar system. So, an understanding of giant planet formation may help us
to determine the occurrence rate of habitable planets.

\end{abstract}



\section{Introduction} \label{sec:intro}

The core accretion model \citep{pol96} of planet formation predicts a 
deficit of planets between the masses of Neptune and Saturn \citep{idalin04,mor09}. 
This desert is a consequence of the runaway accretion of hydrogen and helium gas 
onto protoplanetary cores that have attained a critical mass of $\sim 10\,M_{\oplus}$. 
These cores are preferentially formed outside the snow line. The rapid growth of gaseous 
envelopes around the cores is quenched by either the severe gas depletion throughout 
the planetary disk before the runaway accretion can begin or the growth of 
$\simgt 100\,M_{\oplus}$ gas giants that clear the gaps in the disk. This process is 
expected to produce numerous $\sim 10\,M_{\oplus}$ ``failed gas giant cores", 
as well as gas giants of $\simgt 100 M_{\oplus}$, but few intermediate-mass giant 
planets of $\sim 20$-$80\,\mearth$. In this Letter, we test this scenario with 
gravitational microlensing observations.

A unique strength of the microlensing method is its sensitivity to low-mass 
planets \citep{ben96} in Jupiter-like orbits \citep{gou92}, beyond the snow line. 
Microlensing is most sensitive to planets located at a projected separation similar to the
Einstein radius, which is given by
\begin{equation}
R_{\rm E}=D_{\rm L}\sqrt{{4GM_{\rm L}\over c^2}\left({1\over D_{\rm L}}-{1\over D_{\rm S}}\right)} 
 = 4.04\,{\rm AU} \sqrt{{M_{\rm L}\over\msun} {D_S\over 8\,{\rm kpc}}\, 4x(1-x)}  \ ,
\label{eq:Re}
\end{equation}
where $M_{\rm L}$ is the lens mass, $G$ is the gravitational 
constant, $c$ is the speed of light, $D_{\rm L}$ and $D_{\rm S}$ are 
the lens and source distances, and $x = D_{\rm L}/D_{\rm S}$. For a typical value
of $4x(1-x) = 0.75$ and $D_{\rm S} =8\,$kpc, this gives $R_{\rm E} = 3.5\,{\rm AU} \sqrt{M_{\rm L}/\msun}$,
which is larger than the snow line at $\sim 2.7\,{\rm AU} (M_{\rm L}/\msun)$ for stars of a solar
mass or less.

In this Letter, we compare the cold-planet mass-ratio function derived by 
the MOA collaboration \citep{suz16} (hereafter S16), using 30 microlens planets,
to predictions based on the core accretion theory \citep{pol96}, which was originally constructed 
to study the formation of our own solar system. 
This theory involves many competing physical processes 
including the formation of protoplanetary embryos as progenitor of terrestrial 
planets, gas accretion onto cores of proto-gas giants, the clearing of gaps 
in the protoplanetary disks, and planetary
migration induced by planets' interaction with their natal disks.  None of these processes can be 
calculated with certainty. Nevertheless, it is possible to introduce a set of 
approximations of these processes to simulate the statistical distribution for 
an ensemble of emerging systems from protostellar disks with an assumed 
evolving distribution of H/He gas and planetesimals. 
These constructs generate population syntheses that can be compared with observations.
We compare the microlensing results to the population synthesis models of two
different groups: \citet{idalin04} and the Bern group \citep{mor09}.

\section{Population Synthesis Calculations} \label{sec:pop}

Population synthesis models generate a set of simulated planetary systems 
that have a distribution of planets in the semimajor-axis, planet-mass plane. 
The planet distribution is thought to depend on the host-star mass, and so 
we have run simulations with different host-star masses ranging from 
$0.08$ to $1.0\,M_{\odot}$. We select a range of host-star masses in logarithmically 
uniform bins to span this mass range, with fewer bins for the Bern models, 
because the Bern code uses more CPU time. For each synthesis model, we 
generate planetary systems with and without the planetary migration effect 
because, as discussed below, the models with migration predicted many 
fewer planets than found by microlensing over a large range of mass ratios. 
The details of these population synthesis models are explained below.

\subsection{IL Model} \label{sec:idalin}

The detailed description of the Ida \& Lin (IL) model is written in a series of 
papers \citep{idalin04, idalin04ii, idalin05iii, idalin08iv, idalin08v, idalin10vi, ida13}. 
The synthesis model includes classical models of planetary growth and migration 
from small planetesimals, combining planetesimal accretion, gas accretion onto 
the planet, type I and II migration, as well as planet-planet scattering between 
all planets. Planetary embryos are set with an initial mass of $10^{20}\,{\rm g}$ with orbital 
separations of $\sim 10\,\times$ the Hill radius of the classical isolation mass 
\citep{kok98} at separations of $0.05-20\, {\rm au}$. 
The self-similar disk model is used for disk evolution. 
%\citep{lyn74}. 
The gas surface density at 10 au is 
distributed in a range of $[0.1, 10]$ times of the minimum-mass solar nebula model 
\citep{hay85} with a log-normal distribution. The initial metallicity of the disk is 
distributed in a range of $[-0.2, 0.2]$ dex with a normal distribution. The IL simulations 
were done with nine different host masses given by
$M_h = 10^\gamma M_{\odot}$, where $\gamma = -1.3, -1.15, -1.0, -0.85, -0.7, ..., -0.1$, and the average
weights used for these simulations were 0.067, 0.047, 0.063, 0.077, 0.093, 0.111, 0.135, 0.163, and
0.243, respectively. Exactly 10,000 simulations were run for each mass ratio.

\subsection{Bern Model} \label{sec:bern}

The Bern model is a global planet formation \citep{ali05} and evolution \citep{mor12} 
model that is based on core accretion. It predicts 
the properties of emerging planetary systems like the masses and orbits of 
the planets based directly on the properties of the parent protoplanetary disk 
(such as the disk mass, dust-to-gas ratio, and lifetime). As described in recent 
reviews \citep{benz14,mor15}, the model couples self-consistently several 
standard astrophysical submodels for planetary formation to compute the evolution 
of the gas and planetesimals disks, the accretion of gas and solids by the 
protoplanets, as well as interactions between the protoplanets (gravitational {\it N}-body 
interaction) and between the gas disk and the protoplanets (orbital migration). 
More specifically, the model consists of the following elements:
\begin{enumerate}
\item a protoplanetary gas disk modeled by the numerical solution of the classical 1D viscous 
evolution equation for the gas surface density 
%\citep{lyn74} 
in an axisymmetric constant $\alpha$-viscosity 
%\citep{sha73} 
disk with stellar irradiation \citep{hue05} and photoevaporation \citep{hol94};
\item the protoplanets' accretion rates of solids modeled by a Safronov-type 
%\citep{saf69}  
rate equation 
from planetesimals of a single size in the oligarchic growth regime \citep{ina01};
\item the disk of planetesimals modeled as a surface density with a dynamical state \citep{for13};
\item the protoplanets' gas accretion rate and planetary interior structure obtained from
solving the standard 1D radially symmetric hydrostatic planet interior structure equations;
%\citep{bod86}, 
\item planetary orbital migration modeled as gas disk-driven non-isothermal Type I and Type II 
migration \citep{dit14}; and 
\item dynamical interactions between the protoplanets modeled with the explicit N-body
integrator Mercury \citep{cha99}. 
\end{enumerate}

In population syntheses 
%\citep{mor09} 
models,  the initial conditions of the model, which are the properties of the protoplanetary disks, 
are varied according to observed distribution of protoplanetary disk properties, and 
the global model is run typically several hundred times in order to synthesize 
the predicted population of model planets. 
The Bern group population synthesis calculations are much more computationally 
intensive than the IL calculations, so these were done at host masses of $M_{\rm h} = 10^\gamma\,M_{\odot}$, where 
$\gamma = -0.903, -0.602, -0.301$, and 0, and the average weights for these masses were 
0.268, 0.211, 0.308, and 0.213, respectively. The number of simulations run for each mass ratio 
was 1387, 4805, 4093, and 1392, respectively, for the standard simulations and 
1150, 1717, 1181, and 1918, respectively, for the migration-free simulations.

\section{Cold Exoplanet Mass-ratio Function from Microlensing} \label{sec:q-function}
\label{sec:cold_mrf}

\begin{figure}
\label{fig:sensitivity}
\begin{center}
\includegraphics[width=100mm]{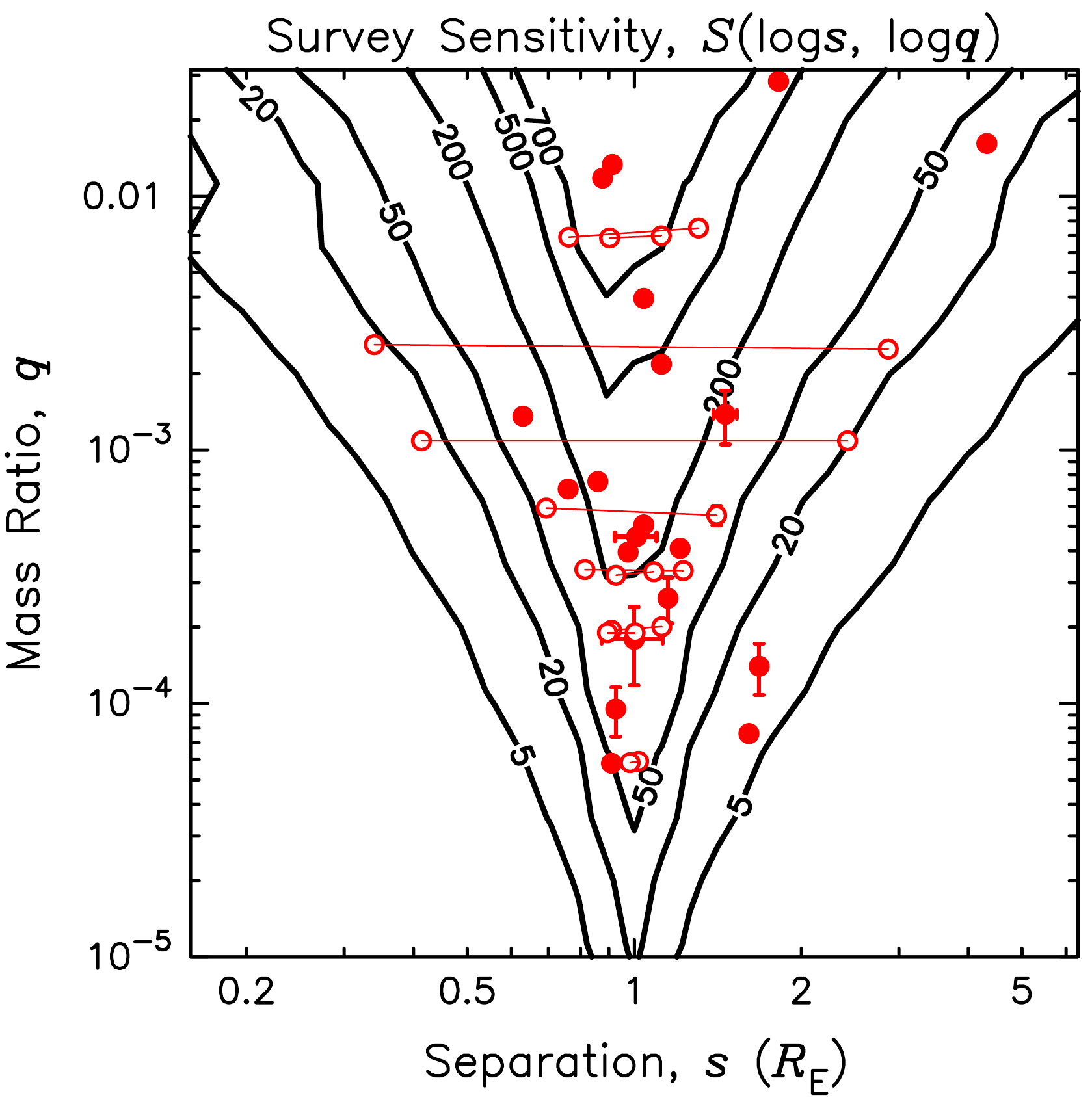}
 \end{center}
\caption{
Exoplanet sensitivity of the combined S16 analysis as a function of mass ratio, $\mathbold{\it q}$, and 
separation, $\mathbold{\it s}$, in Einstein radius units. The contours give the number of planet 
detections expected if each lens system in the combined S16 sample has a planet with the specified 
$q$ and $s$ values, and the red spots indicate the parameters of the planets in the sample. The open 
red spots connected by red indicate high-magnification events with a degeneracy between models
with $s \leftrightarrow 1/s$.\label{fig:sensitivity}
}
\end{figure}
%\clearpage

\begin{figure}
\begin{center}
\includegraphics[width=150mm]{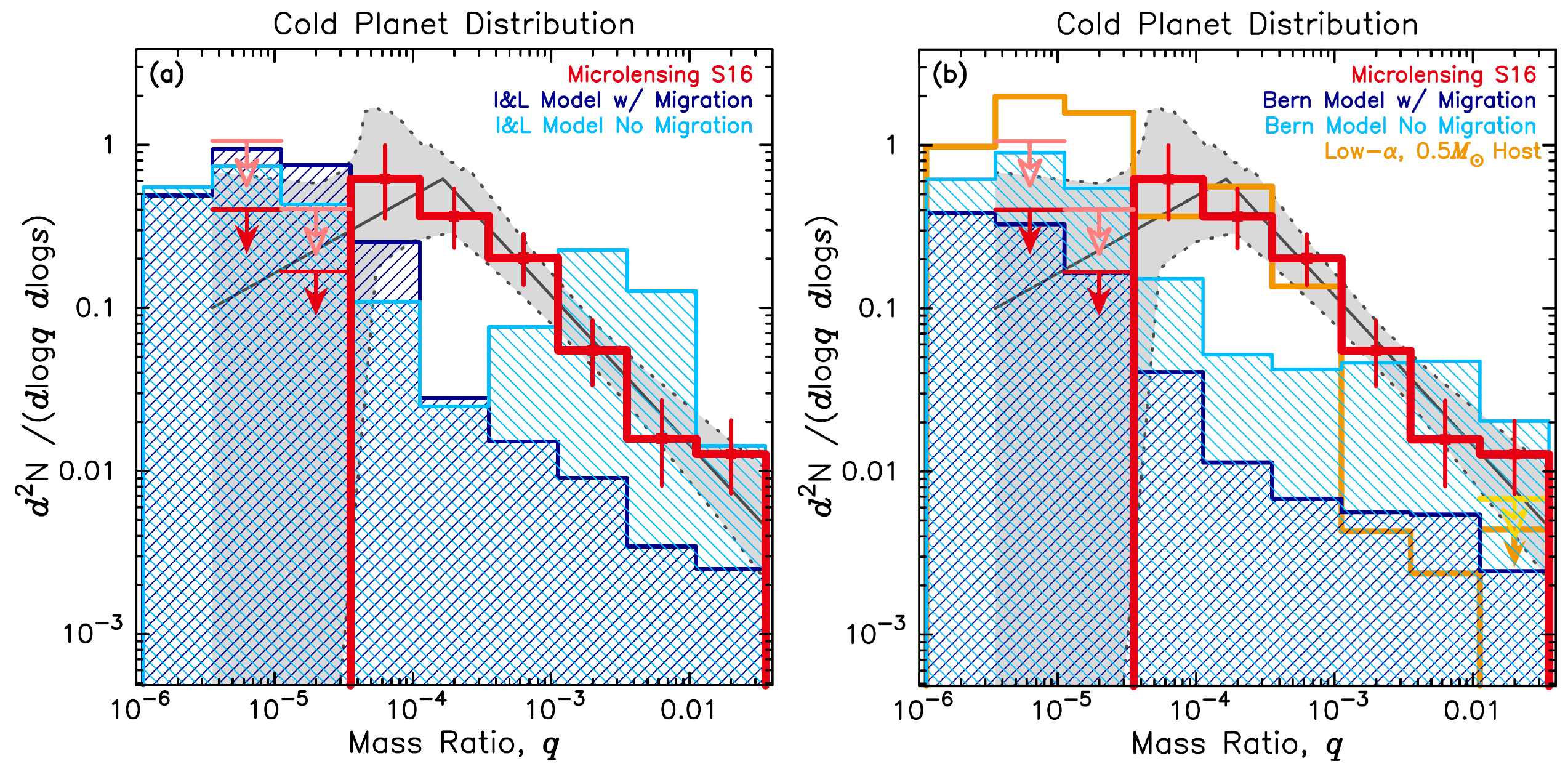} 
\end{center}
\caption{
Planet to host-star mass-ratio function measured by microlensing compared to 
the planet distribution from core accretion theory population synthesis models.
The red histogram shows the measured mass-ratio distribution, with the best-fit broken power-law model
and its $1\sigma$ range indicted by the solid black line and gray shaded regions. The red and pink
arrows indicate the $1\sigma$ and $2\sigma$ upper limits on the mass-ratio bins without planet detections.
The dark and light blue histograms are the predicted mass-ratio functions from the default population
synthesis models with migration, and the alternative migration-free models
from the Ida \& Lin (a) and Bern (b) simulations, respectively. For the Bern model, we also show results for a run with
$2.9\times$ lower disk viscosity for $0.5\,M_\odot$ host stars only as gold histogram in panel (b).
\label{fig:q-func}
}
\end{figure}

Our analysis is based on the S16 statistical analysis \citep{suz16} of the combined 
sample of the 1474 well-characterized microlensing events from the MOA survey 
sample combined with earlier, smaller microlensing samples \citep{gou10,cas12}. 
This sample includes 30 planets with mass ratios, $q$, and separations, $s$, that 
are displayed in Figure \ref{fig:sensitivity} and \ref{fig:q-func}. 
Figure \ref{fig:sensitivity} shows the survey sensitivity for the combined survey, 
which is the sum of the detection efficiencies for all the events in the survey. 
The planets in this S16 sample are indicated by red dots, and planets with 
uncertainty in the separation, $s$, due to the close-wide degeneracy of 
high-magnification events are indicated as open red circles connected by red lines. 
The contours in this figure indicate the number of planets that would be detected 
if each event had a planet with the specified $s$ and $q$ values. The sensitivity to 
planets generally increases at larger $q$ values, but the older microlensing 
samples \citep{gou10,cas12} did not consider planets with mass ratios $q > 0.01$, 
which gives rise to a slight decrease in sensitivity for $q$ values just above 0.01.

%\clearpage

It is this survey sensitivity that is used to convert the microlensing results to 
the power-law mass-ratio functions investigated in S16, but our comparison 
to population synthesis models requires a slightly different approach. The 
population synthesis models provide a set of simulated planetary systems 
for each assumed host-star mass. The planet-star mass-ratio values are 
determined directly from these simulations, but the projected separations, 
$s$ in Einstein radii are not directly produced by the population synthesis 
calculations. We use a standard Galactic model \citep{han95} to produce 
a probability distribution of primary lens masses for each of the 1474 
microlensing events in the S16 sample. Then, for each of these events, 
we run 4000 random trials for the IL and Bern group simulations for each 
of the 1474. In each trial, we randomly select a lens distance and a host-mass 
bin based on the lens mass and distance probability distribution for 
the event under consideration, and we randomly select one of the simulated 
planetary systems from that mass bin. Next, we select a random orientation 
for that planetary system to determine the $s$ value for that event. Finally, 
for the trial for each of the S16 events, we apply the S16 detection efficiency 
as a function of $q$ and $s$ for that event to determine if the simulated 
planets are detected. This is equivalent to simulating 4000 trials of the 
S16 observations, and the total number of events simulated is 
$4000 \times 1474 = 5.9 \times 10^{6}$.

This procedure automatically selects planetary host-star masses from the 
distributions expected for our sample of microlensing events, using our assumed
Galactic model. The microlensing rate imparts a weight that scales as 
$\sqrt{M_{\rm L}}$, and there is an additional weighting from the microlensing
event and planet detection efficiencies. These effects push the median mass of
the host stars produced by our survey up to about $0.6\msun$, as explained in S16.

\section{Predicted Microlens Mass-ratio Function from Population Synthesis}
\label{sec:conversion}

Figure \ref{fig:q-func} compares the S16 results to the population synthesis models 
by IL (a) and the Bern group (b). Both sets of default
models assumed the widely adopted values for the effective viscosities 
($\alpha =10^{-3}$), aspect ratio ($h=H/r$), heavy element abundance, 
mass flux (${\dot M}$), and depletion time scale ($\tau_{\rm dep}$),
% based on the minimum mass nebula model or the observed properties 
for the natal disks of protoplanets \citep{har98}. Despite some 
differences in their prescriptions for the growth of planetesimals 
and gas accretion onto gas giants, the default IL and Bern models 
similarly reproduce the mass-period distribution for gas giants around 
solar-type stars found by the early radial velocity (RV) surveys \citep {cum08}. 
In order to account for a population of short-period gas giants found 
by these observations, these default models include the effect of 
planetary migration \citep{lin86}. 

Comparison between
the population synthesis results and the S16 data shows some significant
differences in the planetary occurrence rate for all mass ratios $q \geq 
10^{-4}$, up to the brown dwarf boundary at $q = 0.03$. This discrepancy 
is a factor of 10--25 at the lower mass-ratios, $10^{-4}\leq q \leq 
10^{-3}$, and it decreases to a factor $\sim$5 for higher mass-ratios, 
$10^{-3}< q \leq 0.03$. One possible culprit for this discrepancy may 
be the loss of such planets due to both type I and II migration.
This conjecture is verified by another series of simulations with identical 
disk-model parameters but without planetary migration. For mass ratios of
$10^{-3}\leq q \leq 10^{-2}$, these migration-free simulations with the Bern
models provide a good match to the S16 data, but the IL models produce an 
overretention of the same population by a factor of $\sim$6.

The migration-free approximation also provides a correction of an 
overestimation of the planetary migration 
%rate \cite{lin86, war97} and reduces 
rate \citep{lin86} and reduces 
an excess of hot Jupiters predicted by the population synthesis models \citep{idalin04}.  
Such a modification may be due to either diffusion across the 
%gap\cite{duffellmacfadyen2013,fung14,kanagawaetal2015},
gap \citep{fung14},
or relative low viscosity \citep{duffell13} (see below). However, these migration-free 
models continue to generate, respectively, 10 and 7 times fewer planets 
with $q=1-4\times 10^{-4}$ than the S16 data.  Figure \ref{fig:q-func} contains four, seven, 
and eight planets in the mass-ratio bin centered at $q = 6.3 \times 10^{-5}$, 
$2.0\times 10^{-4}$, and $6.3\times 10^{-4}$, respectively.  Based on 
their large deficits in Figure \ref{fig:q-func}, we show in Table \ref{tab:poisson} the low 
%Poisson probability for the predicted occurrence rates from the IL and 
%Bern models to match or exceed the detected event rate.
Poisson probability for the predicted number of the planet detections from the IL and 
Bern models to match or exceed the actual number of planet detections in S16.

\begin{table}
\centering
\caption{Poisson Probability for $N\geq N_{\rm obs}$ for Mass-ratio Bins with $ -4.45 < {\rm log} q < -2.95$ \label{tab:poisson}}
\medskip
%\begin{tabular}{rcccccc}
\begin{tabular}{lcccc}
\hline
  & \multicolumn{2}{c}{I \& L model} & \multicolumn{2}{c}{Bern model} \\
 ${\rm log}\, q$  & Migration & No 
 %& optimal 
 & Migration & No 
 \\
 \hline
   $-4.2$ & $0.14$ & $2.0 \times 10^{-2}$ & $1.5 \times 10^{-3}$ & $4.4 \times 10^{-2}$ \\
   $-3.7$ & $1.1 \times 10^{-5}$ & $6.0 \times 10^{-6}$ & $1.7 \times 10^{-8}$ & $2.3 \times 10^{-4}$ \\
   $-3.2$ & $3.6 \times 10^{-6}$ & $2.6 \times 10^{-2}$ & $1.8 \times 10^{-9}$ & $1.3 \times 10^{-3}$ \\
\hline
\end{tabular}
\end{table}

Both the simulated and observed S16 samples obey Poisson statistics, but the statistical uncertainty
on the simulated planets are negligible compared to the Poisson uncertainties for the observed sample.
So, we use the Poisson statistics for the S16 events only to compare the data to the simulations.
These numbers obey Poisson statistics and are used for the computation of
the Poisson probabilities reported in Table \ref{tab:poisson}.  
As mentioned above, Table \ref{tab:poisson} shows the Poisson probabilities for the three $q$ bins centered on
$q = 2.0\times 10^{-4}$ for the standard and migration-free simulations by both groups.
Population synthesis models fail this comparison for the bin centered at $q = 2.0\times 10^{-4}$
for each of the IL and Bern population synthesis models. The two adjacent bins, centered at
$q = 6.3\times 10^{-5}$ and $q = 6.3\times 10^{-4}$ also fail this comparison in almost every case.

Since the median mass of host stars probed by the microlensing survey 
is $M_{\rm L} \approx 0.6\msun$, this mass-ratio desert in the population 
synthesis models corresponds to planets with masses $20-80\mearth$. 
According to the core accretion scenario, solid planetesimals coagulate into
cores that begin to gradually accumulate gaseous envelopes after their mass
%exceeds $\sim 10\mearth$\cite{pol96,lissauer09,raf11}. 
exceeds $\sim 10\mearth$ \citep{pol96}. 
Runaway gas accretion 
is initiated after the envelopes' mass becomes comparable to that of the 
cores. The planets' growth rate accelerates on a time scale inversely 
proportional to their mass until gas is severely reduced either throughout 
the disk or in the proximity of their orbit.  Since the duration of runaway 
growth is generally shorter than the global disk depletion time scale, planets 
form less frequently in the $q$ desert, which is bounded by the 
critical-core mass for the onset of rapid gas accretion and the 
mass for the formation of tidally induced gap \citep{linpap1980}.
Gas giants' asymptotic $q$ is determined by the magnitude of viscosity 
($\alpha$) and aspect ratio ($h$) \citep{linpap93}. For the disk
parameters we adopted in both the default and migration-free models, 
it exceeds $4 \times 10^{-4}$ near the snow line.

\section{Discussion and Conclusions}\label{sec:conclude}

The incompatibility between the default population synthesis models 
and the exoplanet mass-ratio function measured by microlensing suggests 
that some of the assumptions we have adopted for the runaway gas 
accretion scenario might be incorrect or incomplete. We note that a similar
contradiction between the core accretion prediction of a sub-Saturn-mass desert
at a short-period orbit \citep{idalin04ii} is also not seen in the Kepler data  \citep{kepler2018},
but the contradiction seems more significant at separations beyond the show line, where
gas giant formation is thought to occur.
Perhaps there are 
some physical processes that can suppress or quench the gas accretion 
rate.  Planetesimals might be captured by the accreting gaseous envelope 
and heat it up, which would slow the accretion \citep{alibert18}. Gas 
giant planets' asymptotic mass can also be lowered in disk regions with 
%smaller\cite{brydenetal2000, dobbs-dixonetal07, fung14} 
smaller \citep{dob07,fung14} 
$\alpha$ and $h$ 
than the values we have adopted in the default and migration-free models.
During the advanced stages of their viscous evolution, protostellar disks 
may indeed have smaller $h$ values than that for the minimum mass nebula 
model \citep{garaud07}, 
especially around low-mass stars. Growing protoplanets may also undergo type III 
migration \citep{masset06}
or gravitational scattering to regions beyond the snow line where MHD 
turbulence may be suppressed by magnetic diffusivity \citep{bai17} to reduce
the magnitude of effective $\alpha$ (below $10^{-4}$).  These regions 
may be manifested in the form of narrow and axisymmetric rings and gaps 
in protostellar disks commonly found by ALMA and they may provide nests 
for the low-mass protoplanets.  These effects need to be explored further
elsewhere.    

The Bern group has performed some preliminary simulations for 
low-viscosity disks (with one-third the value of $\alpha$ as the default
models) around $0.5M_\odot$ host stars. In addition, the tidal interaction of protoplanets with the 
protoplanetary gas disk can lead to gap opening, which in turn reduces the planet's gas 
accretion rate by reducing the 
surface density of the gas in the vicinity of the planet \citep{dan10}. 
These results (represented
by the gold histogram in Figure \ref{fig:q-func}b) yield compatibility with the 
microlensing data in the $10^{-4} < q <10^{-3}$ mass-ratio range. 
Lowering of protoplanets' tidal truncation mass also leads to the
formation of too many planets with $q < 3\times 10^{-5}$ such that 
center of the $q$ desert is shifted to lower $q$ values with less 
paucity. However, these model parameters also suppress the emergence 
of planets with $q > 10^{-3}$. These issues with the low-$\alpha$ 
models are verified by analogous IL simulations that also fail to 
reproduce the population of Jupiter-mass planets found by the RV 
surveys around solar-type stars.
However, this difficulty might be mitigated if some higher mass-ratio planets could be formed
by mechanism other than core accretion, such as gravitational instability in the 
protoplanetary disk \citep{dur07,forgan13,boss17}.

%We have also done some preliminary calculations with the Bern model of low $\alpha$-viscosity disk 
%to allow gap opening to slow the runaway gas accretion phase, using an $\alpha = 7 \times 10^{-4}$ value, 
%which is $\sim$3 times smaller the value, $\alpha = 2 \times 10^{-3}$ used for the simulations shown 
%in Figure \ref{fig:q-func}. In addition, it is considered that the tidal interaction of protoplanets with the 
%protoplanetary gas disk leads to gap opening, which in turn reduces the planet's gas accretion rate by reducing the 
%surface density of the gas in the vicinity of the planet \citep{dan10}.

Since the microlensing and RV surveys sample host stars with somewhat
different, but overlapping, mass distributions, a stellar-mass-dependent $q$ distribution 
(due to variations in the values of $\alpha$, $h$, and 
$\tau_{\rm dep}$) remains a possibility.  In principle, gaps in 
the $q$ distribution may be smoothed out when the survey 
samples include a range of stellar masses.  
%If so, it may be only the systems that undergo runaway gas accretion 
%that are habitable, as it has been suggested that the delivery of water 
%to the inner planets of our Solar System may be a consequence of 
%the runaway gas accretion process \citep{ray17}.
We plan to test 
this possibility with high angular resolution follow-up 
observations of the S16 planetary microlensing events using 
adaptive optics (AO) observations on the {\it Keck} telescopes 
under a recently approved NASA Key Strategic Mission Support 
(KSMS) program and {\it Hubble Space Telescope} observations \citep{bha18}.
When the host star is detected, its mass can usually be determined 
from a measurement of the host-star brightness combined with constraints 
from the microlensing light curve. In these follow-up observations, 
it is important to measure the separation of the lens and source 
stars \citep{bat15,ben15} to confirm the lens star identification 
because confusion with other stars, like a binary companion to 
the lens or source, is possible \citep{bha17,kos17}. 

One of the first targets observed under the {\it Keck} KSMS program is the S16
event OGLE-2012-BLG-0950, which has been found \citep{bha18} to have a host 
mass of $M_h = 0.58\pm 0.04\,M_{\odot}$ and a planet mass of $39 \pm 
8\,M_\oplus$. This planet is right in the middle of the mass gap expected 
from the runaway gas accretion process in the default and migration-free models. 
There are also several solar-type stars outside the S16 sample that have planets 
in this intermediate-mass range, including 
the microlens planet OGLE-2012-BLG-0026Lc \citep{bea16}, with 
$M =46.0 \pm 2.6\,M_\oplus$, and two 
planets from the HARPS survey \citep{mayor11} just inside the snow 
line with $M \sin i \sim 50\,M_\oplus$.  If the mass measurements of 
the other planets in the S16 sample reveal a number of other planets 
in this predicted mass gap, this would rule out the host-mass dependence 
as the reason for the discrepancy between the microlensing data and 
the expectations for the runaway gas accretion process. 
The resolution of this discrepancy may help us to understand habitability of inner planets because 
it has been suggested that the delivery of water to the inner planets of our solar system may be a consequence of the runaway gas accretion process \citep{ray17}.
Ultimately, 
the exoplanet survey \citep{ben02} of the Wide Field Infrared Survey 
Telescope \citep{penny18} will perform a much more comprehensive survey that 
will probe planets beyond the orbital separation of Venus with sensitivity 
down to the mass of Mars ($0.1\,M_\oplus$).

\acknowledgments

D.P.B., A.B., and C.R.  were supported by NASA through grant NASA-80NSSC18K0274. 
 The MOA project is supported by JSPS KAKENHI grant Nos. JSPS24253004, 
 JSPS26247023, JSPS23340064, JSPS15H00781, JP16H06287 and JSPS17H02871.
 The work by C.R. was supported by an appointment to the NASA Postdoctoral Program at the 
 Goddard Space Flight Center, administered by USRA through a contract with NASA.
 C.M. acknowledges the support from the Swiss National Science Foundation under grant 
 BSSGI0\_155816, ``PlanetsInTime." Parts of this work have been carried out within the frame of the 
 National Center for Competence in Research PlanetS supported by the SNSF.
 Work by N.K. was supported by JSPS KAKENHI grant No. JP18J00897.
 N.J.R. is a Royal Society of New Zealand Rutherford Discovery Fellow.

\end{document}